\documentclass[aps,prl,twocolumn,superscriptaddress,groupedaddress]{revtex4}
\usepackage{graphicx}
\usepackage{color}
\usepackage{bm}       
\usepackage{amssymb} 
\usepackage{amsmath}
\usepackage{amsfonts}
\usepackage{amsthm}

\begin{document}

\title{Elemental chalcogens as a minimal model for chiral charge and orbital order}

\author{Ana Silva}
\author{Jans Henke}
\author{Jasper van Wezel}
\affiliation{Institute for Theoretical Physics, Institute of Physics, University of Amsterdam, 1090 GL Amsterdam, The Netherlands}

\date{\today}

\begin{abstract}
Helices of increased electron density can spontaneously form in materials containing multiple, interacting density waves. Although a macroscopic order parameter theory describing this behaviour has been proposed and experimentally tested, a detailed microscopic understanding of spiral electronic order in any particular material is still lacking. Here, we present the elemental chalcogens Selenium and Tellurium as model materials for the development of chiral charge and orbital order. We formulate minimal models capturing the formation of spiral structures both in terms of a macroscopic Landau theory and a microscopic Hamiltonian. Both reproduce the known chiral crystal structure and are consistent with its observed thermal evolution and behaviour under applied pressure. The combination of microscopic and macroscopic frameworks allows us to distil the essential ingredients in the emergence of helical charge order, and may serve as a guide to understanding spontaneous chirality both in other specific materials and throughout materials classes.
\end{abstract}

\maketitle

The bulk transition metal dichalcogenide \emph{1T}-TiSe$_2$ has been shown, uniquely, to harbour a charge density wave transition that breaks inversion symmetry in a chiral way~\cite{ishioka,vanwezellittlewood,vanwezel,castellan,iavarone}. The spontaneous formation of helicity is well-known in magnetic materials, in which spins may wind around a propagation direction to yield spirals of magnetisation. In contrast, helices of increased electronic density necessarily require the onset of charge order to be accompanied by a simultaneous onset of orbital order~\cite{vanwezel,fukutome,fukutome2}. Although this restricts the class of materials in which chiral charge order may appear~\cite{prerequisites}, it has nevertheless been theoretically suggested to play an important role in determining material properties of various transition metal dichalcogenides~\cite{prerequisites,TaS2spain,TaS2vanwezel}, and even cuprate high-temperature superconductors~\cite{Raghu,Raghu2,Gradhand}.

Chiral charge order was suggested to be present in \emph{1T}-TiSe$_2$ based on indirect experimental evidence~\cite{ishioka}. In addition, several predictions arising from a theoretical model of chiral charge order, based on a Ginzburg-Landau free energy expansion, have been experimentally confirmed~\cite{vanwezel,castellan,iavarone}. Nevertheless, it has proven difficult to obtain direct experimental confirmation of the broken inversion symmetry. The main reason for this is believed to be the presence of small, nanometer wide, domains of varying handedness~\cite{iavarone}, which are averaged over by almost all direct bulk probes. A microscopic understanding of the chiral phase transition, going beyond the predictions of macroscopic order parameter theory, is thus essential for guiding further experiments into this novel type of charge and orbital order.
\begin{figure}
\begin{centering}
\includegraphics[width=\columnwidth]{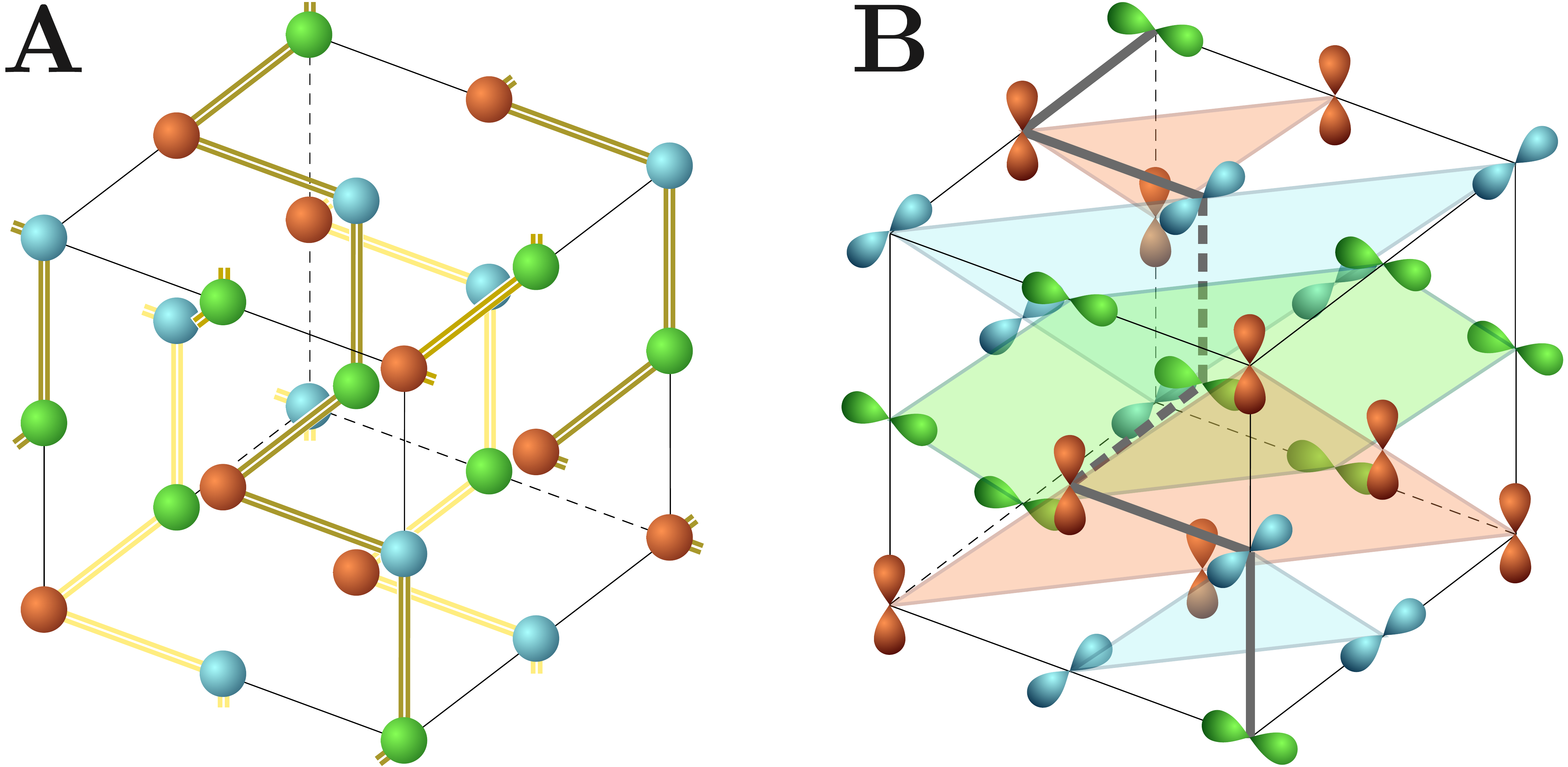}
\end{centering}
\caption{\label{fig:Xtal}Chiral charge and orbital order in elemental chalcogens. {\bf a)} The chiral crystal structure of Se and Te can be understood as a spiral arrangement of short bonds in a simple cubic parent lattice. All atoms in the crystal are of the same type. Different colours indicate the three possible local configurations of short bonds, and the chiral unit cell includes one atom of each color. {\bf b)} Because the short bonds involve charge transfer between specific orbitals only, the chiral crystal lattice is also orbital ordered. Indicated here are the least occupied orbitals. The shaded planes connect like orbitals and are included as a guide to the eye. They are perpendicular to the spiral axis of the crystal structure.}
\end{figure}

Despite its relatively simple crystal structure, the ordered state of \emph{1T}-TiSe$_2$ involves too many orbitals and electronic bands for the construction of a microscopic theory to be a straightforward exercise, or to lead to intuitive insight into the mechanism underlying the formation of chiral charge order. Here, we therefore take an alternative approach, and construct a minimal microscopic model for the appearance of spiral chains in the atomic structure of the elemental chalcogens Se and Te. These materials do not exhibit a charge ordering transition at any temperature, but their atomic lattices are well known to be chiral at ambient conditions. The handedness of a given sample of Te or Se can be straightforwardly determined by measuring either its diffraction pattern or its optical activity~\cite{tanaka}. The crystal structure of Se and Te can be understood as consisting of short bonds arranged along helices in a simple cubic parent structure, as shown schematically in Fig.~\ref{fig:Xtal}a~\cite{tanaka}. This picture can in fact be taken literally, and the spiral bond order can be shown to be an instability of a hypothetic parent phase with simple cubic lattice structure~\cite{fukutome,fukutome2,vanwezellittlewood}. The charge ordering transition leading from the simple cubic to the chiral phase is of the same type as the chiral transition in \emph{1T}-TiSe$_2$~\cite{vanwezel}, but as we show here, it can be understood on the level of an explicit microscopic model, explaining how different types of electron-phonon coupling and Coulomb interactions conspire to form the spiral structure. The minimal model constructed in this work is thus presented as a prototype description for the formation of chiral charge and orbital order in general.

~\\
{\bf Intuitive picture.} Before presenting both a macroscopic Ginzburg-Landau and a microscopic mean-field theory for the formation of chiral charge order in elemental chalcogens, we will first give an intuitive picture showcasing their basic ingredients. The starting point for this is a simple cubic lattice structure. Both Se and Te actually possess the chiral crystal structure shown in Fig.~\ref{fig:Xtal}a for any temperature at ambient pressure. Upon melting Te however, short-ranged chiral order in the fluid phase has been found to disappear, and a homogeneous metallic phase to form instead, not much above the melting point~\cite{fluid1,fluid2}. This observation can be understood as a latent structural phase transition in the crystal lattice of Te, which is preempted by the material melting before the transition temperature can be reached. The crystal structure of the hypothetical high-temperature phase can in that case be assumed to be simple cubic, since the element Po, which sits just below Te in the periodic table and thus has the same configuration of valence electrons, crystallises into a simple cubic rather than a chiral structure~\cite{polonium}. The expected phase transition into a chiral orbital ordered phase in Po is suppressed by the presence of strong spin-orbit coupling~\cite{koreans}, which allows the parent simple cubic structure to emerge.

Elemental chalcogens have four valence electrons in their outermost $p$-shell ($2/3$ filling). Within a simple cubic lattice potential the $p_x$, $p_y$, and $p_z$ orbitals are degenerate, and may be chosen to point along the crystallographic $x$, $y$, and $z$ axes respectively. Since their wave functions are elongated in a single direction, the overlap between neighbouring $p_x$ orbitals on the $x$ axis will be larger than that between neighbouring $p_x$ orbitals on the $y$ or $z$ axes. Taking this difference to the extreme limit, we will consider a minimal model in which the overlap between orbitals aligned in a head-to-toe manner is non-zero, and all other overlaps are neglected. Although quantitatively unrealistic, this assumption does not change the qualitative physics of the chiral phase transition, and naturally leads to an intuitive minimal model.

In a tight-binding model starting from the simplified overlap integrals, an electron in a $p_i$ orbital, with $i \in \{x,y,z\}$, can only hop in the $i$ direction, onto a neighbouring $p_i$ orbital. The simple cubic lattice is thus filled with independent one-dimensional chains of $p_x$ orbitals, interwoven by similar one-dimensional chains in the $y$ and $z$ directions. As there is no inter-chain hopping in any direction, the electronic structure consists of three one-dimensional bands, oriented along three orthogonal directions. Within the cubic first Brillouin zone each 1D band introduces a pair of parallel planar Fermi surfaces, whose intersections again form a cube. The Fermi surface is extremely well-nested, and a charge density wave instability is thus expected to emerge~\cite{Peierls}. In fact, a single nesting vector ${\bf Q}$, corresponding to a body diagonal of the cube of intersecting Fermi surfaces, connects each point on any of the Fermi surface sheets to a point within a parallel sheet. The dominant instability will therefore be towards the formation of charge density waves $\rho_j({\bf x}) = \rho_0 + A \cos({\bf Q} \cdot {\bf x})$ in each of the three orbital sectors (labeled by $j$), who all share the same propagation direction ${\bf Q}$. Here $\rho_0$ is the average charge density in the normal state, and $A$ is the amplitude of the charge modulation.

The presence of a non-zero electron-phonon coupling causes the atomic lattice to deform in response to the charge modulations. The resulting displacement waves ${\bf u}_j({\bf x}) = \tilde{u} {\bf e}_j \sin({\bf Q} \cdot {\bf x})$ have the same wave vector as the charge modulations, but a polarization ${\bf e}_j$ whose direction is determined by the anisotropy of the local electron-phonon coupling matrix elements~\cite{vanwezel}. In a chain of $p_x$ orbitals without any overlap in directions other than $x$, the electron-phonon coupling is maximally anisotropic, and the displacement direction ${\bf e}$ will be purely along $x$. Within the simple cubic lattice, each atom is affected simultaneously by three displacement waves, corresponding to the charge density waves in the three $p$-orbitals on the atom. The actual displacement is then simply the sum of the three orthogonal components ${\bf u}_j$. 

The charge density wave in each orbital chain can be shifted along its propagation direction by the addition of a phase: $\rho_j({\bf x}) \propto \cos({\bf Q} \cdot {\bf x} + \varphi_j)$. The vector ${\bf Q}$ shows the charge order in Se and Te to have period three in all directions. The coupling of charge and lattice modulations then restricts the phase $\varphi_j$ to be a multiple of $\pi/3$, so that the point of highest charge always sits either on an atomic site or bond (resulting in a site-centered charge density wave, or a bond-centered charge density wave~\cite{vandenbrink}). If additionally we consider a Coulomb interaction between the charges in orthogonal orbitals on the same site, the charge maxima along one chain will prefer to avoid the charge maxima along other chains. This effectively couples the phases in different orbital sectors, so that a configuration with $\varphi_i - \varphi_{i+1}  = \pm 2\pi/3$ becomes energetically favourable. The final configuration with three orthogonal charge density waves shifted with respect to each other produces precisely the charge redistribution and lattice deformations shown in Fig.~\ref{fig:Xtal}a, which agree with the experimentally observed crystal structure of Se and Te~\cite{fukutome,vanwezellittlewood}.

Notice that because the charge maxima in orthogonal orbital chains avoid one another, each atom in the final structure ends up with a single $p$ orbital being more occupied than the other ones. The chiral charge ordered structure is therefore also automatically an orbital ordered phase, as shown in Fig.~\ref{fig:Xtal}b. The handedness of the crystal structure can thus be equally interpreted as signifying the rotation of the least occupied orbital wave function upon traversing the crystal.

{\bf Macroscopic order parameter theory.} To understand the emergence of chirality from the interplay between three order parameters in different orbital sectors in more detail, we first construct a Landau free energy describing the transition. The dimensionless order parameters $\alpha_j( {\bf x}) $ represent the (periodic) modulation of the average charge density within a given chain:$\rho_j({\bf x}) = \rho_0 ( 1 + \alpha_j({\bf x}))$. Starting from three non-interacting orbital sectors,  the first contribution to the free energy is just the sum of the Landau free energies $F_j$ for three independent charge density waves, with $F_j= \int d^3 x \; a({\bf x}) \alpha_j^2 + b({\bf x}) \alpha_j^3 + c({\bf x}) \alpha_j^4$. Notice that the presence of a discrete lattice is taken into account by expanding the coefficients in the free energy in terms of the reciprocal lattice vectors, so that for example $a({\bf x}) = a_0 + a_1 \sum_n e^{i {\bf G}_n \cdot {\bf x}}+ \dots$~\cite{mcmillan}. Terms with $n>0$ originate from the electron-phonon coupling in a more microscopic model. In the following, we take into account the first order contributions only.

The on-site Coulomb interaction between electrons in orthogonal orbitals provides the additional interaction terms $F_{\text{Coul}} = \sum_{j} \int d^3 x \; A_0 \alpha_j \alpha_{j+1}$, which couple the three order parameters. The periodic charge distributions can be written as $\alpha_j( {\bf x}) = \psi_0 \cos( {\bf Q} \cdot {\bf x} + \varphi_j )$, with the amplitude $\psi_0$ equal for all three order parameters, and $\varphi_j$ the spatial shift of the charge density wave along $j$ with respect to the atomic lattice. Performing the spatial integration over positions ${\bf x}$ in the expression for the full free energy $F$, results in an expression that depends both on the amplitude and phases of the order parameters:
\begin{align}
F  =& \frac{3}{2} a_0 \psi_0^2 + \frac{9}{8} c_0 \psi_0^4  + \frac{1}{4} b_1 \psi_0^3  \sum_j \cos(3\varphi_j) \notag \\ 
& + \frac{1}{2} A_0 \psi_0^2 \sum_{j} \cos(\varphi_j - \varphi_{j+1}).
\label{GL}
\end{align}
As usual, the temperature dependence of the quadratic term proportional to $a_0$ determines when $\psi_0$ first obtains a non-zero value, and charge order sets in. The combination of the final two terms, arising from the electron-phonon coupling and the Coulomb interaction respectively, determine the values of the phases $\varphi_j$ in the presence of a non-zero order parameter. They can be simultaneously minimised by first of all taking $\varphi_1=n \pi/3$, where $n$ is an odd or even integer depending on the sign of $b_1$. Physically, this difference corresponds to the charge order being either site or bond centered. Additionally, the relative phase differences should be chosen as $\varphi_j - \varphi_{j+1} = \pm 2\pi/3$. These solutions are then precisely the left and right handed chiral configurations consisting of mutually shifted one-dimensional charge density waves discussed in the previous section, one of which is shown in Fig.~\ref{fig:Xtal}a.

It is instructive to compare the free energy of Eq.~\eqref{GL} to the one given in Ref.~\onlinecite{vanwezel}, describing the charge ordered phase in \emph{1T}-TiSe$_2$. The higher complexity of the atomic configuration in TiSe$_2$, as compared to the pure elements Se and Te, results in three charge density waves along different propagation directions, as well as a difference in strength of the electron-phonon coupling between Ti and Se sites. Nevertheless, the route to chiral charge and orbital order is largely the same as that observed in Eq.~\eqref{GL} for Se and Te. The onset of charge order is determined by terms that do not involve the phases of the individual charge density wave components. Instead, a term arising from the electron-phonon coupling favours values of the phases in individual sectors to be such that charge maxima fall on top of atomic sites or bonds. The on-site Coulomb interaction finally, provides a coupling between charge density waves in different orbital sectors. The coupling may be indirect in the case of bond-centered charge order, like in TiSe$_2$, where the variation of bond densities implies charge redistributions on the atomic sites, which are then subject to local Coulomb interactions. The coupling between orbital sectors leads to relative phase shifts between them, and hence the emergence of a chiral charge and orbital ordered pattern.

{\bf Applied pressure.} Upon the application of large uniform pressure, Se and Te undergo a series of structural transitions into non-chiral phases~\cite{pressure1,pressure2}. Within the minimal model considered here, the suppression of chirality under pressure can be captured by introducing a pressure dependence of the critical temperature. Using the values for the phases appropriate in the chiral state, the quadratic coefficient in the free energy then becomes:
\begin{align}
a_0(T,P) = \frac{3 b_1^2}{32 c_0} + \frac{A_0}{2} + \alpha \left[ \frac{T}{T_C^0} + \frac{P}{P_C^0} - 1\right]
\label{a0TP}
\end{align}
Here $P_C^0$ is the critical pressure at zero temperature, while $T_C^0$ is the critical temperature at zero applied pressure. Notice that for the purposes of this minimal model, the relation between critical temperature and pressure is assumed to be linear, and that the high-pressure, non-chiral phase can only be simple cubic in structure. In spite of these simplifications, the free energy expansion captures the suppression of chirality by pressure, and may be straightforwardly extended to qualitatively examine the result of applying for example uniaxial rather than uniform pressure.
\begin{figure}
\begin{centering}
\includegraphics[width=\columnwidth]{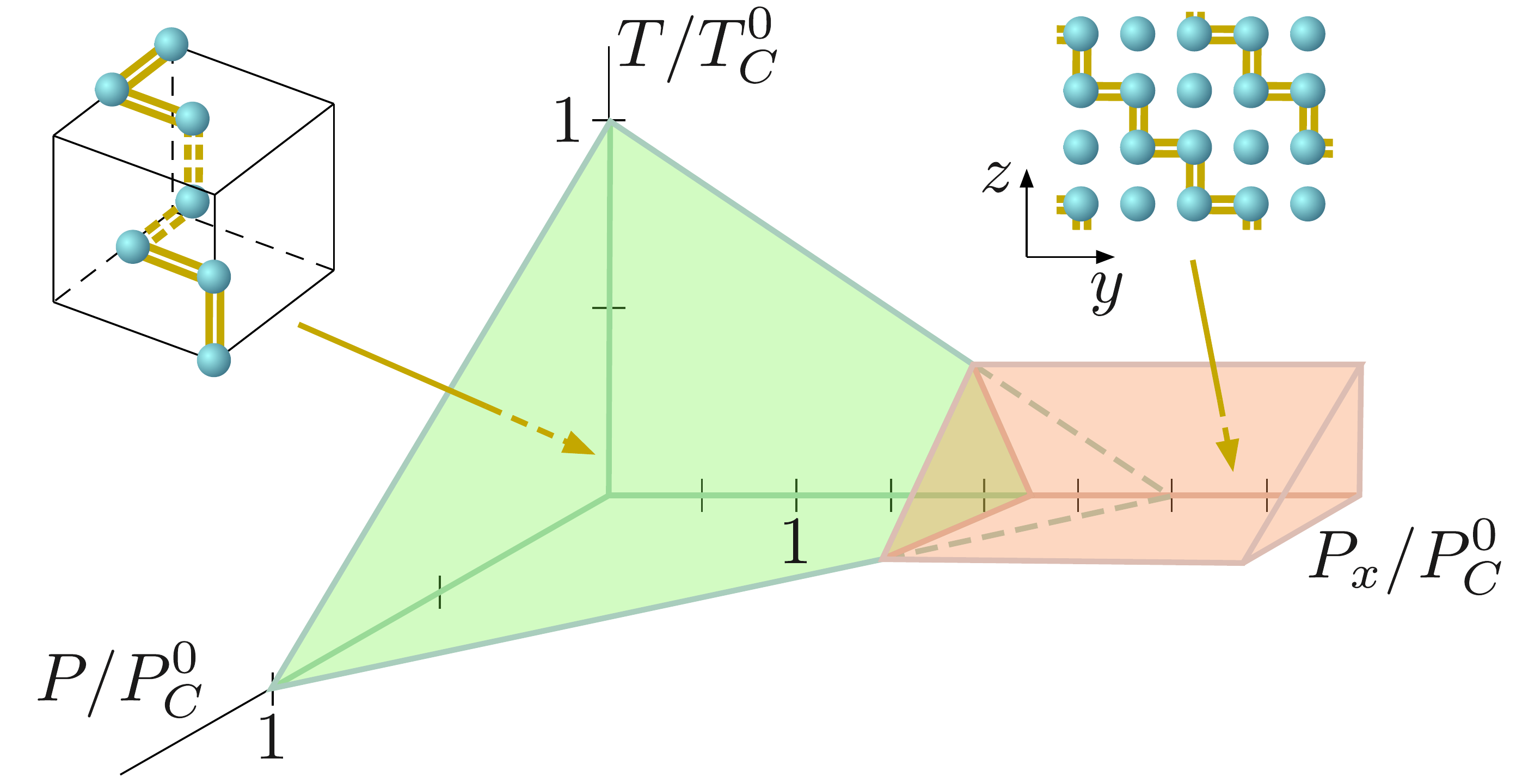}
\end{centering}
\caption{\label{fig:zigzag} Schematic phase diagram indicating the relative stability of the chiral and zig-zag phases within the free energy of Eq.~\eqref{GL_P}. The zig-zag phase consists of planes  perpendicular to the direction of applied uniaxial strain with zig-zag chains of short bonds, as indicated in the inset. The applied pressure along the $x$ axis is parameterised as $P+P_x$, with $P$ the uniform pressure and $P_x$ the uniaxial strain component. The applied pressure along the $y$ and $z$ axes is simply $P$. Notice that the linearity of the critical lines and planes in this phase diagram is a direct consequence of the simplified thermal and pressure dependences assumed in Eq.~\eqref{a0TP}.}
\end{figure}

Anisotropic phases can be included in the minimal model by allowing the amplitudes $\psi_j$ of charge density waves in different orbital sectors to develop independently, and for each to have its own critical temperature, depending on the amount of pressure applied along its particular axis. The expression for the free energy then becomes:
\begin{align}
F  =& \sum_j \frac{1}{2}  a_{0}(T,P_j) \psi_j^2 + \frac{3}{8} c_0 \psi_j^4 + \frac{1}{4} b_1 \psi_j^3  \cos(3\varphi_j)  \notag \\ 
& ~~~~~~ + \frac{1}{2} A_0 \psi_j \psi_{j+1} \cos(\varphi_j - \varphi_{j+1}).
\label{GL_P}
\end{align}
Applying pressure along a single axis only, the charge order in a singe direction may be suppressed without destroying the order in orthogonal directions. The result is a phase of stacked planes each containing zigzag charge order, as indicated schematically in Fig.~\ref{fig:zigzag}, along with the phase diagram resulting from this minimal model. The anisotropic structure resulting from the minimal model agrees both with the predictions of an earlier semiclassical approach in terms of so-called vector charge density waves~\cite{fukutome}, and with the experimental observation of layered structures in Se under high pressures~\cite{pressure1,pressure2}.

{\bf Microscopic model.} To see how the terms in the Landau free energy emerge from the interplay of microscopic degrees of freedom, we construct a minimal Hamiltonian model for Se and Te. The starting point is again a two-third filled $p$-shell within the simple cubic lattice. We then include a tight-binding approximation for the bare electronic band structure, an on-site Coulomb interaction, and the influence of lattice distortions on both the kinetic and potential energies of electrons. 

Including hopping only between neighbouring orbitals that are aligned head-to-toe in the simple cubic lattice, the tight-binding part of the Hamiltonian can be written as $H_{\text{TB}} = t  \sum_{{\bf x},j} \hat{c}_{j}^{\dagger}({\bf x}) \hat{c}_{j}^{\phantom \dagger}({\bf x} + {\bf a}_j) + \text{H.c.}$, where $\hat{c}_{j}^{\dagger}({\bf x})$ creates an electron in orbital $j$ on position ${\bf x}$, and ${\bf a}_j$ is the simple cubic lattice vector in direction $j$. The overlap integral $t$ is positive because the overlapping orbital lobes on neighbouring sites have opposite signs. The Coulomb interaction acts on-site, through the interaction $H_{\text{Coul}} = V \sum_{{\bf x},j} \hat{c}_{j}^{\dagger}({\bf x}) \hat{c}_{j}^{\phantom \dagger}({\bf x}) \hat{c}_{j+1}^{\dagger}({\bf x}) \hat{c}_{j+1}^{\phantom \dagger}({\bf x})$. The displacement $\hat{u}_{j}^{\phantom \dagger}({\bf x})$ of the atom on position ${\bf x}$ in the direction of $j$ may be written in terms of the phonon operator $\hat{b}_{j}^{\dagger}({\bf x})$, which is taken to be a dispersionless Einstein mode with $H_{\text{boson}} = \hbar \omega \sum_{{\bf q},j} \hat{b}_{j}^{\dagger}({\bf q}) \hat{b}_{j}^{\phantom \dagger}({\bf q})$. The electron-phonon coupling consists of two terms:
\begin{align}
 \hat{H}^{(1)}_{\text{e-ph}} &=  \alpha^{(1)} \sum_{{\bf x},j} \left(\hat{u}_{j}^{\phantom \dagger}({\bf x}) - \hat{u}_{j}^{\phantom \dagger}({\bf x} + {\bf a}_j) \right) \cdot \notag \\
&~~~~~~~~~~~~~ \left( \hat{c}_{j}^{\dagger}({\bf x}) \hat{c}_{j}^{\phantom \dagger}({\bf x} + {\bf a}_j) + \hat{c}_{j}^{\dagger}({\bf x} + {\bf a}_j) \hat{c}_{j}^{\phantom \dagger}({\bf x}) \right) \notag \\
\hat{H}^{(2)}_{\text{e-ph}} &=  \alpha^{(2)}  \sum_{{\bf x},j} \left( \hat{u}_{j}^{\phantom \dagger}({\bf x} + {\bf a}_j) - \hat{u}_{j}^{\phantom \dagger}({\bf x} - {\bf a}_j) \right) \hat{c}_{j}^{\dagger}({\bf x}) \hat{c}_{j}^{\phantom \dagger}({\bf x}).
\label{eph}
\end{align}
The first type of electron-phonon coupling represents the change of electronic kinetic energy with varying bond length. If an interatomic distance is decreased, the orbital overlap across the affected bond increases. The second process reflects the change of electronic potential energy with a variation in local ionic density. If a given atom is approached more closely by its two neighbours, the potential energy of electrons located on the central position is lowered to compensate for the larger density of positive core charges. 

The full Hamiltonian can be diagonalised in the mean field approximation by introducing Ansatz averages reflecting the possible ordered states found in the Landau free energy analysis:
\begin{align}
\langle \hat{c}_{j}^{\dagger}({\bf x}) \hat{c}_{j}^{\phantom \dagger}({\bf x}) \rangle &= \rho_0 + A \cos\left( {\bf Q} \cdot {\bf x} + \varphi_j \right) \notag \\
\langle \hat{c}_{j}^{\dagger}({\bf x}) \hat{c}_{j}^{\phantom \dagger}({\bf x} + {\bf a}_j) \rangle &= \sigma_0 + B \cos\left(  {\bf Q} \cdot ({\bf x} + {\bf a}_j)/2 + \chi_j \right) \notag \\
\langle \hat{u}_{j}^{\phantom \dagger}({\bf x}) \rangle &= \tilde{u} \sin\left(  {\bf Q} \cdot {\bf x} + \phi_j \right).
\end{align}
Here $A$ is the mean field expectation value describing modulations of the on-site charge density, $B$ represents the bond-density variations, and $\tilde{u}$ measures the variations in atomic positions. The mean fields $A$ and $B$ can be directly related to the macroscopic order parameter $\alpha$ appearing in the Landau free energy expansions above. Using these definitions, the Hamiltonian decomposes into a fermionic and a bosonic part. The latter can be straightforwardly diagonalised by introducing shifted boson operators~\cite{wezel10_TiSe2_CDW}, and relates the atomic displacements to the electronic order parameters by setting $\tilde{u} = {2\sqrt{3}}/{\hbar \omega} (2B\alpha^{(1)} e^{(\chi_j-\phi_j)} -A \alpha^{(2)} e^{i(\varphi_j-\phi_j)})$. Demanding the displacement to be real restricts the difference between the phases of any two order parameters to be an integer multiple of $\pi$. 

The fermionic part of the mean field Hamiltonian can be diagonalised numerically, and the ground state values of the phases and order parameters determined self-consistently. In the presence of electron-phonon coupling, but with no on-site Coulomb interaction, the three orbital sectors are independent from one another and each develops an individual charge density wave. The phases are simply $\varphi_j = n_j {\pi}/{3}$, with $n_j$ integer, which includes both non-chiral solutions in which the $n_j$ are all equal, and chiral ones. For any non-zero value of the Coulomb interaction this degeneracy is lifted, and the left- and right-handed chiral charge ordered configurations with $n_{j}-n_{j+1} = \pm {2\pi}/{3}$ become the lowest energy states. 

\begin{figure}
\begin{centering}
\includegraphics[width=\columnwidth]{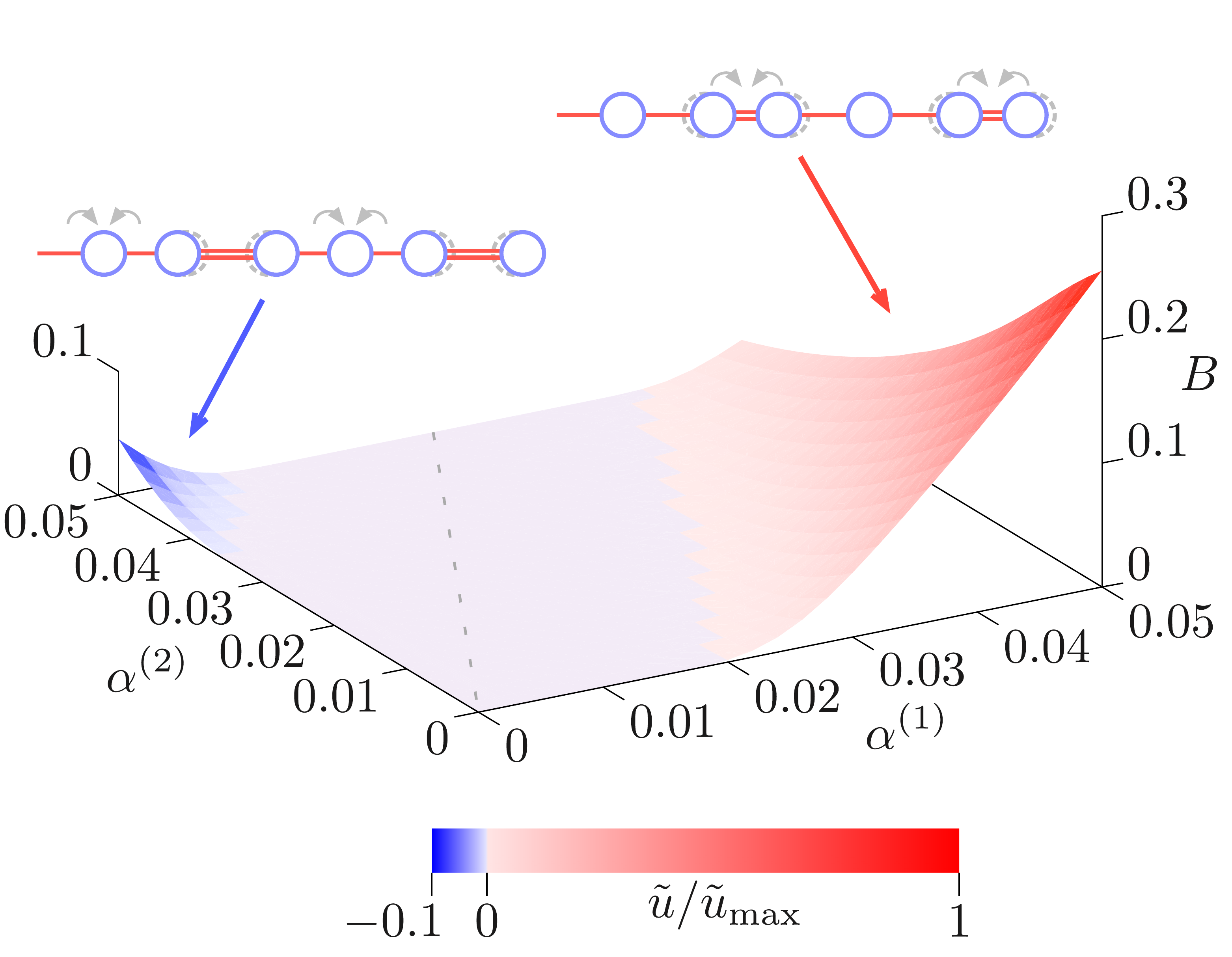}
\end{centering}
\caption{\label{fig:phasediagram}The ground state phase diagram as a function of the two contributions to the electron-phonon coupling in Eq.~\eqref{eph}. The vertical axis measures the charge density wave order parameter $B$, while the colouring indicates the normalised atomic displacement. Each  kind of electron-phonon coupling favours a particular type of chiral charge and orbital ordered state, both of which can be constructed from one-dimensional chains with relative phase shifts. The two kinds of chains appropriate for the two types of electron-phonon coupling are shown schematically in the region where they dominate. Increased bond density is indicated by double lines, while increased on-site electronic density results from charge transfer along the curved arrows.}
\end{figure}

For each handedness, the $n_j$ may be odd or even multiples of ${\pi}/{3}$. These solutions correspond to the location of maximum charge in each charge density wave being either bond-centered or site-centered, as indicated in the insets of Fig.~\ref{fig:phasediagram}. Which of these phases has the lowest energy depends on the balance between the different types of electron-phonon coupling. As shown in the phase diagram of Fig.~\ref{fig:phasediagram}, the bond-centered solution dominates for large $\alpha^{(1)}$, while the site-centered one is consistent with large $\alpha^{(2)}$. The atomic structure observed in elemental Se and Te corresponds to the bond-centered charge order~\cite{tanaka}, with $\alpha^{(1)}$ prevailing. 

Within the chiral phase, the short bonds in the three orbital chain directions connect to form a spiral. The resulting enlarged unit cell and atomic structure agree with the experimentally observed structure for Te and Se, shown schematically in Fig.~\ref{fig:Xtal}a. The displacements in the $x$, $y$ and $z$ directions arise from charge order in chains of $p_x$, $p_y$, and $p_z$ orbitals respectively. The modulation of charge density can thus also be seen as a spatial modulation of orbital occupation. Because the charge density waves in the three orthogonal directions are shifted by $2{\pi}/{3}$ with respect to each other, each site in the atomic lattice of Se and Te has precisely one less-occupied $p$-orbital next to two others that remain equally occupied. Drawing only the least occupied orbitals results in Fig.~\ref{fig:Xtal}b, which clearly shows that the chiral charge ordered state is also an orbital ordered state. The handedness of the orbital order is the same as that of the structural order, and can be seen for example by following the rotation of the least occupied orbital as one progresses through the crystal along the ordering ${\bf Q}$ vector (a body diagonal of the original simple cubic structure). The emergence of orbital order in conjunction with chiral charge order is inevitable, since both arise from the same relative phase shifts between charge density waves in distinct orbital sectors.

~\\
{\bf Discussion.} Indirect evidence for the emergence of chiral charge and orbital order has been found in the low-temperature phase of the layered transition metal dichalcogenide \emph{1T}-TiSe$_2$~\cite{ishioka,vanwezel,castellan}. In addition, experimental predictions from a macroscopic Landau theory for the chiral state in this material were successfully tested~\cite{castellan}. The broken inversion symmetry, however, has not been observed directly by any experiment yet. Probing the bulk helicity is likely complicated by the presence of small domains of opposite handedness, of which indirect signatures are seen in scanning-tunneling microscopy experiments~\cite{iavarone}. In addition, the interplay between many different orbitals located throughout the chiral unit cell prevent overly simplified theoretical models from being applicable and complicate the extraction of physical insight from realistic microscopic models~\cite{monney,zhu_tbp}.

Having an alternative model material, which harbours a similar chiral state but is structurally simple and well-understood, is therefore crucial to aid in building a general understanding of the novel charge and orbital ordered phase. Here, we argue that the elemental chalcogens Tellurium and Selenium constitute precisely such model materials. We construct minimal models for these materials capturing the essential ingredients in the formation of their chiral structures, both in terms of a macroscopic Landau theory and a microscopic mean-field description. 

Comparing the results presented here to the chiral phase of \emph{1T}-TiSe$_2$, highlights both the universal and material-specific aspects of inversion symmetry breaking through combined charge and orbital order. In both cases, the starting point is a material with multiple density wave instabilities in its electronic structure, residing in distinct orbital sectors. The different orbital orientations lead to differently polarised displacement waves in both materials. The on-site Coulomb repulsion then causes maxima of different density waves to repel each other. This results in shifts or relative phase differences between the density waves, which break inversion symmetry and yield the known chiral crystal structure. Because the density waves originate in distinct orbital sectors, the relative phase differences imply simultaneous charge and orbital order. 

The driving mechanism underlying the density wave instabilities in \emph{1T}-TiSe$_2$ likely differs from that in the elemental chalcogens~\cite{wezel10_TiSe2_CDW,abbamonte}, but plays no role in determining whether or not the combined state will be chiral. In contrast to the chalcogens, the propagation vectors for different density waves in TiSe$_2$ are all different. The order is also site-centered in Se and Te, but bond-centered in TiSe$_2$. Finally, the site-centered Coulomb repulsion providing the coupling between different density waves, yields an indirect interaction between the bond-centered charges in the case of TiSe$_2$. 

Whether or not the phase shifts induced by local Coulomb interactions break inversion symmetry, depends sensitively on the crystal structure in which they reside~\cite{prerequisites,TaS2vanwezel}. If the crystal structure in the presence of the phase shifts includes a mirror symmetry, the result is not chiral even if inversion symmetry is broken. The mechanism described above then instead causes the formation of a polar charge and orbital ordered state, as seen for example in \emph{2H}-TaS$_2$~\cite{TaS2spain,TaS2vanwezel}.  In the case of TiSe$_2$ as well as Se and Te however, the crystal symmetries favour the formation of a chiral charge and orbital ordered state.

The theoretical understanding developed here, of how chiral charge and orbital order emerges in elemental Se and Te, can be used as a guiding principle for the understanding of similar phases in other materials. These may include other elements and transition metal dichalcogenides, but the simplicity of the minimal models presented in this work suggests the main mechanism to be  be applicable generically to materials harbouring multiple simultaneous density wave instabilities. As long as charge order develops in distinct orbital sectors that are coupled by a local interaction, relative phase shifts will occur and generically lead to the spontaneous breakdown of inversion symmetry.

~\\ \indent
{\bf Acknowledgments} \\
J.v.W. acknowledges support from a VIDI grant financed by the Netherlands Organisation for Scientific Research (NWO).


\end{document}